\documentclass[a4paper]{jpconf}
\usepackage{graphicx}
\usepackage[square,sort&compress]{natbib}

\def\aap{A\&A} 
\def\apj{ApJ} 
\def\apjl{ApJL} 
\def\mnras{MNRAS} 
\def\nat{Nature} 
\def\physrep{Phys.~Rep.} 

\def\eg{{e.g.}} 

\def\s{{\rm s}} 
\def\ms{{\rm m}\s} 

\def\muas{\mu{\rm as}} 

\begin{document}

\title{Testing General Relativity with High-Resolution Imaging of Sgr A*}

\author{Avery E. Broderick$^1$, Abraham Loeb$^2$}

\address{$^1$,$^2$ Institute for Theory and Computation, Harvard-Smithsonian
  Center for Astrophysics, Cambridge, MA, 02145, USA}

\ead{abroderick@cfa.harvard.edu, aloeb@cfa.harvard.edu}

\begin{abstract}
Submilliarcsecond astrometry and imaging of the black hole Sgr A* at
the Galactic Center may become possible in the near future at infrared
and submillimetre wavelengths.  This resolution is sufficient to
observe the silhouette the supermassive black hole in the Galactic
center casts upon background emission.  However, more exciting is the
prospect of observing ``hot spots'' in the accretion flow.  Here we
discuss how such measurements may be used to test not only the
consistency of General Relativity, but also the validity of the
Kerr metric in particular.
\end{abstract}

\section{Introduction}
Testing strong field gravity remains one of the primary objectives of
observational astronomy.  Due to their compact nature, black holes
provide an ideal environment to do this.  Nevertheless, an
unambiguous confirmation of strong field relativity has been elusive
thus far.

There have been a number of attempts to probe the strong gravity
regime, including observations of the relativistically broadened Fe
K$\alpha$ line
(see, \eg, \cite{Pari-Brom-Mill:01,Reyn-Nowa:03}), interpretations of
quasi-periodic oscillations (QPOs)
(see, \eg, \cite{Remi:05,Genz_etal:03}), and multiwavelength spectropolarimetric
observations
(see, \eg, \cite{Conn-Star-Pira:80,Laor-Netz-Pira:90,Brod-Loeb:06}).
However, the interpretation of each of these are dependent upon
unknown accretion physics, making the implications for general
relativity ambiguous.

For example, the failure to find an expected correlation between
the variability in the Fe K$\alpha$ line emission and the soft X-ray continuum
implies that the simplest emission models for the Fe K$\alpha$
observations are incomplete
(see, \eg, \cite{Wang_etal:99,Chia_etal:00,Lee_etal:00,Wang-Wang-Zhou:01,Weav-Gelb-Yaqo:01}),
though attempts to rectify this with the inclusion of strong
gravitational lensing have been made \cite{Matt-Fabi-Reyn:97}.
In addition, alternative explanations for the formation of the broad
iron lines exist (see, \eg, \cite{Elvi:00,You-Liu-Chen-Chen-Zhan:03}),
further complicating their interpretation.

The most commonly discussed black hole QPO's are those observed in the
X-ray spectra of stellar-mass black hole candidates.  These typically
have $Q$'s of $10$--$100$ and are at kilohertz frequencies.  Unfortunately,
in the absence of a definitive theory for how these are produced, the
identification of these with the epicycles of black hole spacetimes is
tenuous at best.  A second class of QPO's are those observed in
supermassive black holes.  In the context of Sgr A*, these are
observed in the near-infrared (NIR) and X-ray bands, have $Q$'s of
$3$, (though see \cite{Bela_etal:06}), and periods on the order of 20
minutes.  If these are interpreted as the Keplerian orbital periods
of the innermost stable circular orbit (ISCO), they imply black hole
spins as high as $0.5$.

More recently, arguments based upon the lack of a thermal peak in the
spectra have been used to infer the absence of a surface in
stellar-mass black hole candidates (see, \eg, \cite{Garcia_etal:01})
and Sgr A* \cite{Brod-Nara:06}.  In the latter case, where the
putative thermal emission due to the small accretion rate peaks in the
near infrared (NIR), this result appears especially robust.  However,
these arguments present only a qualitative confirmation of the
observational consistency of general relativity.

In contrast, it is now technologically feasible to image a black hole
directly.  A background illuminated black hole will appear in
silhouette, with an angular size of roughly twice that of the horizon
\cite{Bard:73}, and may be directly observed.  With an expected
resolution of $\sim20\,\muas$, submillimeter very-long baseline
interferometry (VLBI) would be able to image the silhouette cast upon
the accretion flow by Sgr A* (with an angular scale of
$\sim50\,\muas$), and M87 ($\sim25\muas$)
\cite{Falc-Meli-Agol:00,Brod-Loeb:05b}.  In principle, detailed
measurements of the size and shape of the silhouette could yield
information about the mass and spin of the central black hole.  In
practice, the interpretation of such an image will likely depend upon
the accretion flow model employed (this is discussed in more detail in
\S\ref{S}).

Sgr A* has exhibited strong flares in the NIR and X-ray
\cite{Ghez_etal:04,Ecka_etal:04,Genz_etal:03,Baga_etal:01}, and more
recently in the submillimeter
\cite{Marr:06}, implying that
the innermost portions of the emitting region are strongly variable.
A simple model for the flares, motivated by the evidence of
periodicity, is that of transient orbiting bright regions, hot spots,
which dominate the flaring luminosity.  Such hot spots appear
inevitable, the product of shocks and magnetic reconnection
events within the accretion flow.  Due to its dynamical and compact
nature, images of such a spot will contain significantly
more information about the spacetime.  Each imaged spot will allow the
measurement of both, the mass and spin of the black hole.  Combining
observations of many hot spots orbiting at different radii provides a
way to test not only the consistency of general relativity, but the
validity of the Kerr metric generally, and the no-hair theorems
specifically.

In \S\ref{S} and \S\ref{HS} we discuss the expected images for a radiatively
inefficient accretion flow (RIAF) and hot spots, respectively, in the context of the
Galactic center.  Some concluding remarks are in \S\ref{C}.

\section{Silhouettes} \label{S}
For a truly back-lit black hole, strong gravitational lensing produces
a silhouette with radius $\sqrt{27} GM/c^2$, roughly twice the size of
the horizon of a non-rotating black hole.  The size and shape are
nearly independent of the black hole spin, though the position of the
silhouette relative to the black hole changes substantially
(see, \eg, \cite{taka:04}).

\begin{figure}[t]
\begin{minipage}{0.5\columnwidth}
\includegraphics[width=\columnwidth]{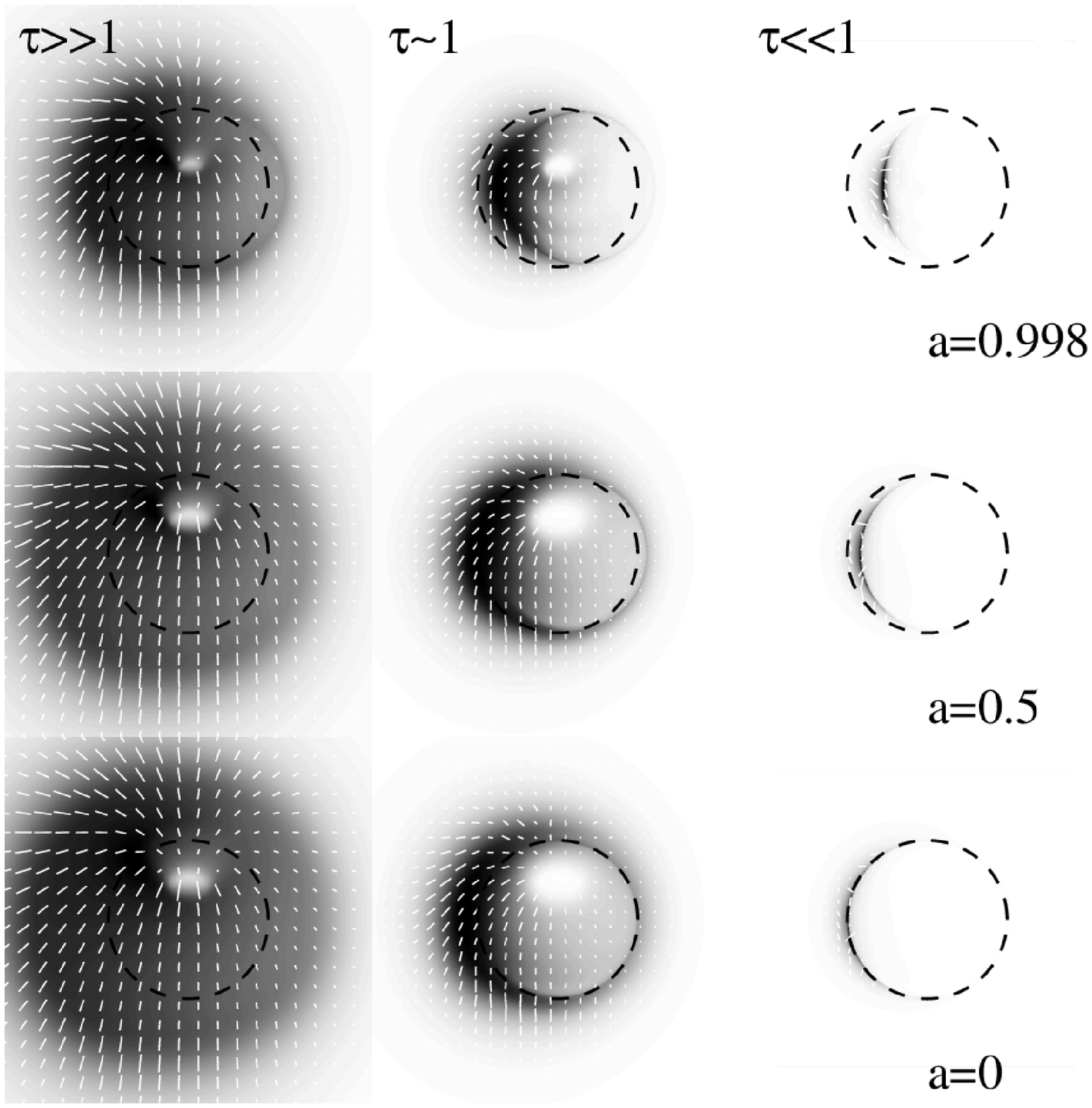}
\caption{\label{fig:silhouettes}The expected silhouettes for low,
  moderate, and high spin (bottom to top) and high, unity, and low
  optical depth (left to right).  Each plot is normalized separately,
  with black being the highest intensity and white corresponding to
  zero.  Overlayed upon the image are white polarization ticks, the
  magnitudes of which corresponds to size of the polarized intensity.
  For reference, the silhouette for the non-rotating black hole is
  shown by the dashed black line.  Note that the hole in the center of
  the optically thick images is due to an evacuated funnel in the
  accretion flow, {\em not} the black hole.  In each plot}
\end{minipage}
\hspace{0.05\columnwidth}
\begin{minipage}{0.5\columnwidth}
\includegraphics[width=\columnwidth]{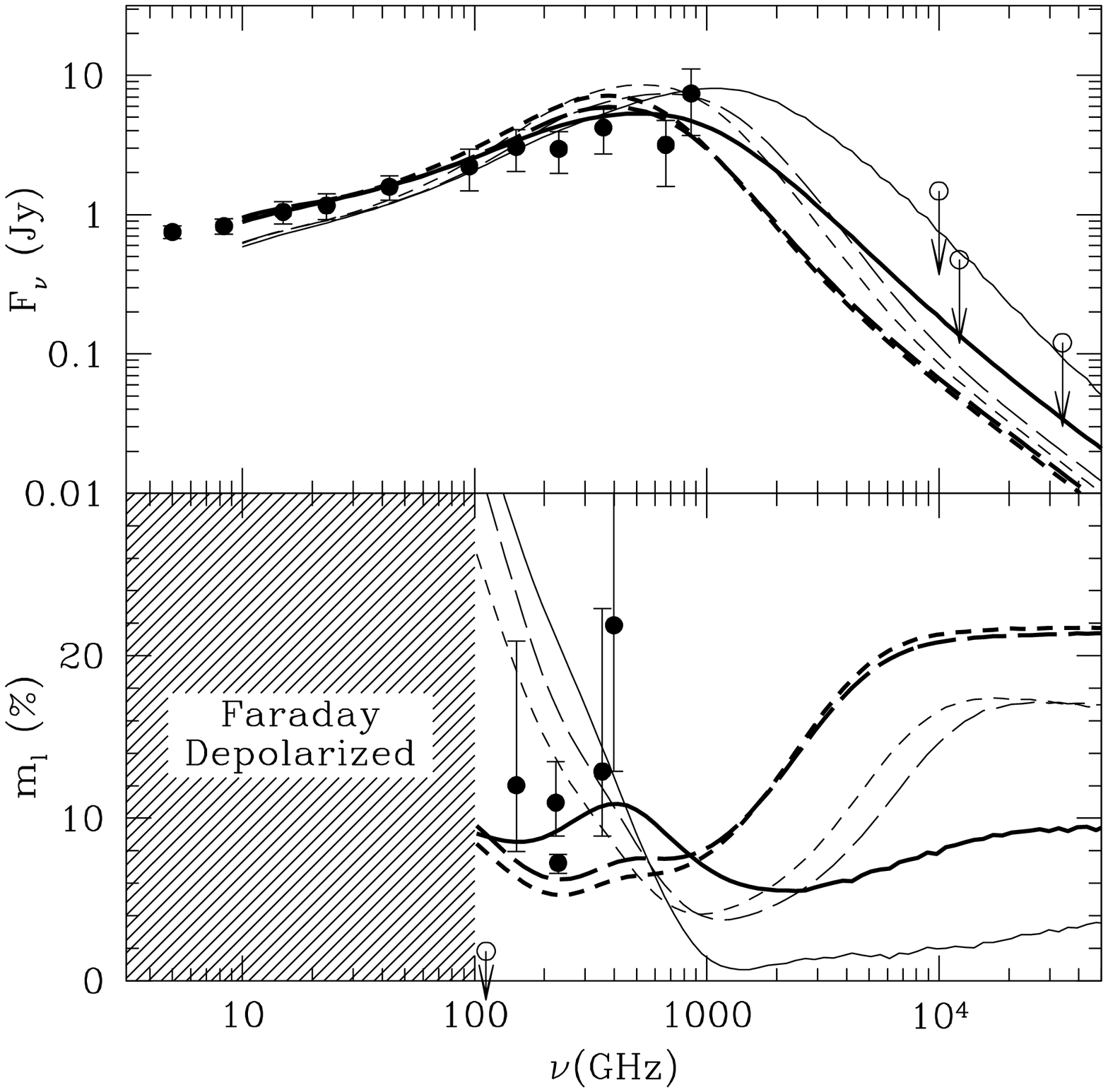}
\caption{\label{fig:centroid_polarization}The total and polarization
  fraction spectra for the non-rotating (short-dash),
  moderately-rotating (long dash) and maximally rotating (solid) black
  holes.  The thick and thin lines correspond to inclination angles of
  $45^\circ$ and $0^\circ$, respectively.  The data are taken from
  \cite{Aitk_etal:00,Bowe-Falc-Saul-Back:02,Yuan-Quat-Nara:04}.}
\end{minipage} 
\end{figure}

However, for a black hole embedded in an accretion flow the silhouette
will generally be asymmetric regardless of black hole spin.  Even in
an optically thin accretion flow asymmetry will result from
special-relativistic effects (aberration and Doppler shifting).  The
degree of asymmetry will be strongly dependent upon the spectral
index, vanishing for $\alpha=-2$ (i.e., a thermally emitting, {\em
  optically thin} disk).  For $\alpha\simeq0$, the asymmetry can be
quite large, as seen in left column of images in Figure
\ref{fig:silhouettes}.

When the opacity of the disk cannot be ignored, as is likely to be the
case in Sgr A* at submillimeter wavelengths, this to will alter the
shape and position of the silhouette.  Indeed, as can be seen in the
left two columns of Figure \ref{fig:silhouettes}, the shape of the
photosphere itself is asymmetric, again due to special relativity.
Hence, it appears unlikely that it will be possible to completely
disentangle the effects of strong gravity from the accretion disk
physics.  Nonetheless, comparisons of the image centroid positions
between the optically thick (where the photosphere is large, and thus
unaffected by the spin of the black hole) and thin limits (where the
emission is dominated by the inner disk edge, presumably at the ISCO)
provides a rough measure of the black hole spin, independent of the
details of the accretion flow \cite{Brod-Loeb:05b}.

A second discriminant between low and high spin may be found in the
net polarization.  In the optically thin limit the emission is
restricted to a small arc, and thus the polarimetric properties of the
source are dominated by the magnetic field in that region.
Conversely, as the optical depth increases the emission is more
distributed and thus the net polarization is an average over a wide
variety of magnetic field directions (compounded with position angle
rotations due to both, the relativistic motion of the disk, and
gravitational lensing).  As a result, quite generally, the polarization
fraction will increase with increasing frequency, plateauing at a
value set by the optically thin spectral index and the spin of the
black hole.  Low asymptotic polarization fractions are then indicative
of high spin, as a consequence of frame dragging.  This is shown for
two magnetic field geometries in Figure
\ref{fig:centroid_polarization}.

Unfortunately, due to the underlying uncertainty in the accretion
model, it will be difficult to make precise measurements of the black
hole spin with these observations.  Alternatively, these measurements
will provide invaluable information on the nature of the accretion
flows in low-luminosity active galactic nuclei.

\section{Hot Spots} \label{HS}
Observations of Sgr A* show strong variability, with the possibility
of periodicity, thus suggesting hot spots in the accretion flow.  This
is not unexpected theoretically as strong inhomogeneities develop in
the innermost regions of general-relativistic magnetohydrodynamic
simulations (see, \eg, \cite{DeVi-Hawl-Krol:03}), although mapping
these into inhomogeneities in the emission is non-trivial.  This is
likely to happen via particle acceleration at strong shocks and/or
magnetic reconnection events.  Both mechanisms will produce a
distribution of compact, non-thermally emitting regions\footnote{Note that
due to the high radiative efficiency of non-thermal electrons in this
environment, this is unlikely to have a significant effect upon the
accretion flow structure.}.  

There are a number of advantages to imaging these ``hot spots''
as opposed to the underlying quiescent accretion flow.  Firstly, due
to their compact, and essentially local nature, they are likely to be
considerably simpler to model than the global structure of the
accretion flow.  Hence, the primary difficulty in using the black hole
silhouettes to measure the mass and spin is naturally solved in this
case.  Secondly, encoded in the images of the hot spot is
significantly more information.  This is because at each instant, the
light rays which connect the spot to the observer fill, and thus are
sensitive to, a far smaller region of the spacetime.  Therefore, as
the hot spot orbits the black hole, the observed image is continually
sampling new regions of the spacetime.  Observing multiple hot
spots at different orbital radii completes the sampling of the spacetime.

\begin{figure}[t]
\begin{minipage}{0.5\columnwidth}
\includegraphics[width=\columnwidth]{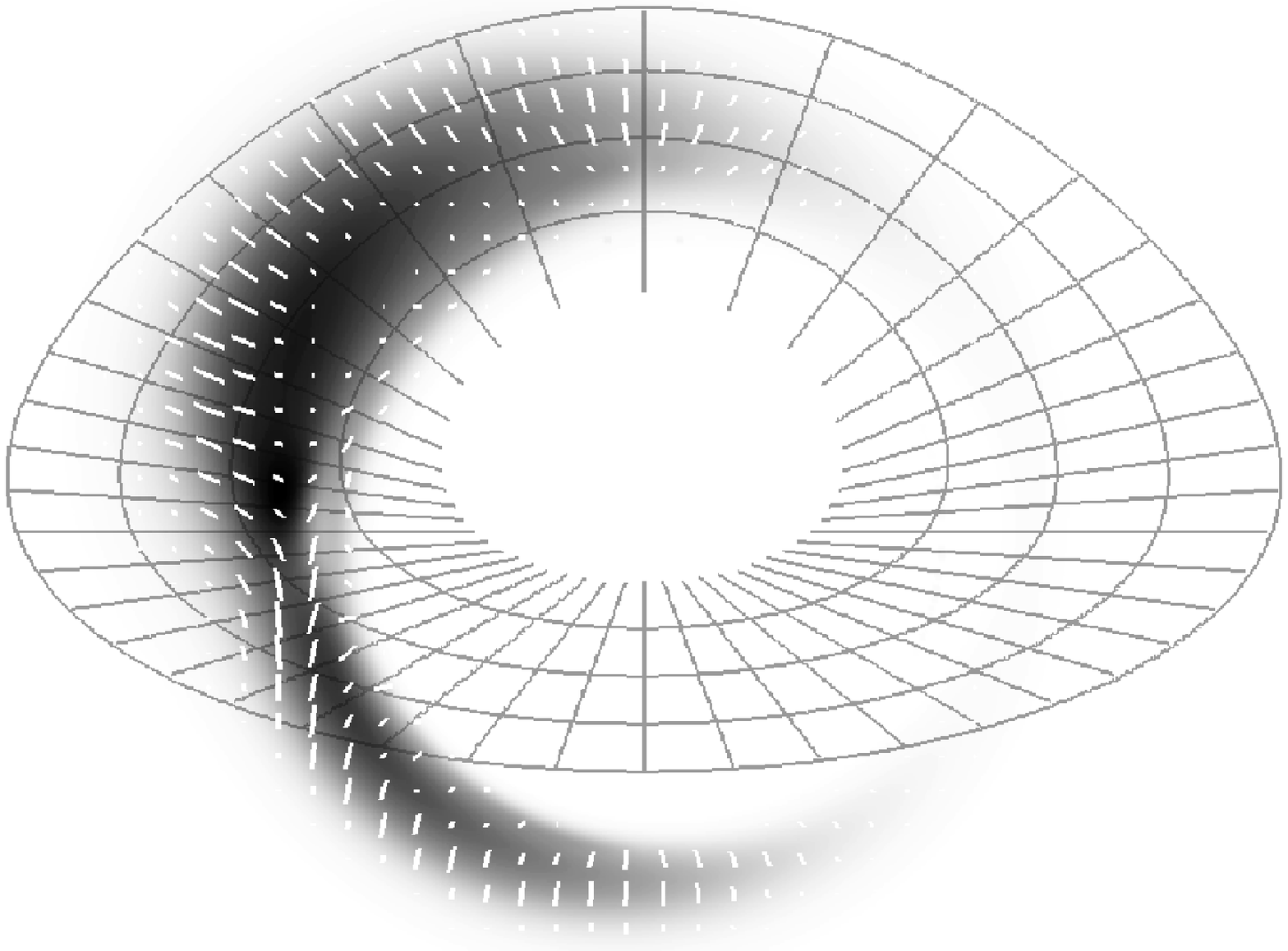}
\caption{\label{fig:spot_image}An instantaneous image of a hot spot
  orbiting a non-rotating black hole at the ISCO.  Overlayed upon the
  intensity are polarization ticks in white.  The lensed upper surface
  of the equatorial plane is shown for reference.}
\end{minipage}
\hspace{0.05\columnwidth}
\begin{minipage}{0.5\columnwidth}
\includegraphics[width=\columnwidth]{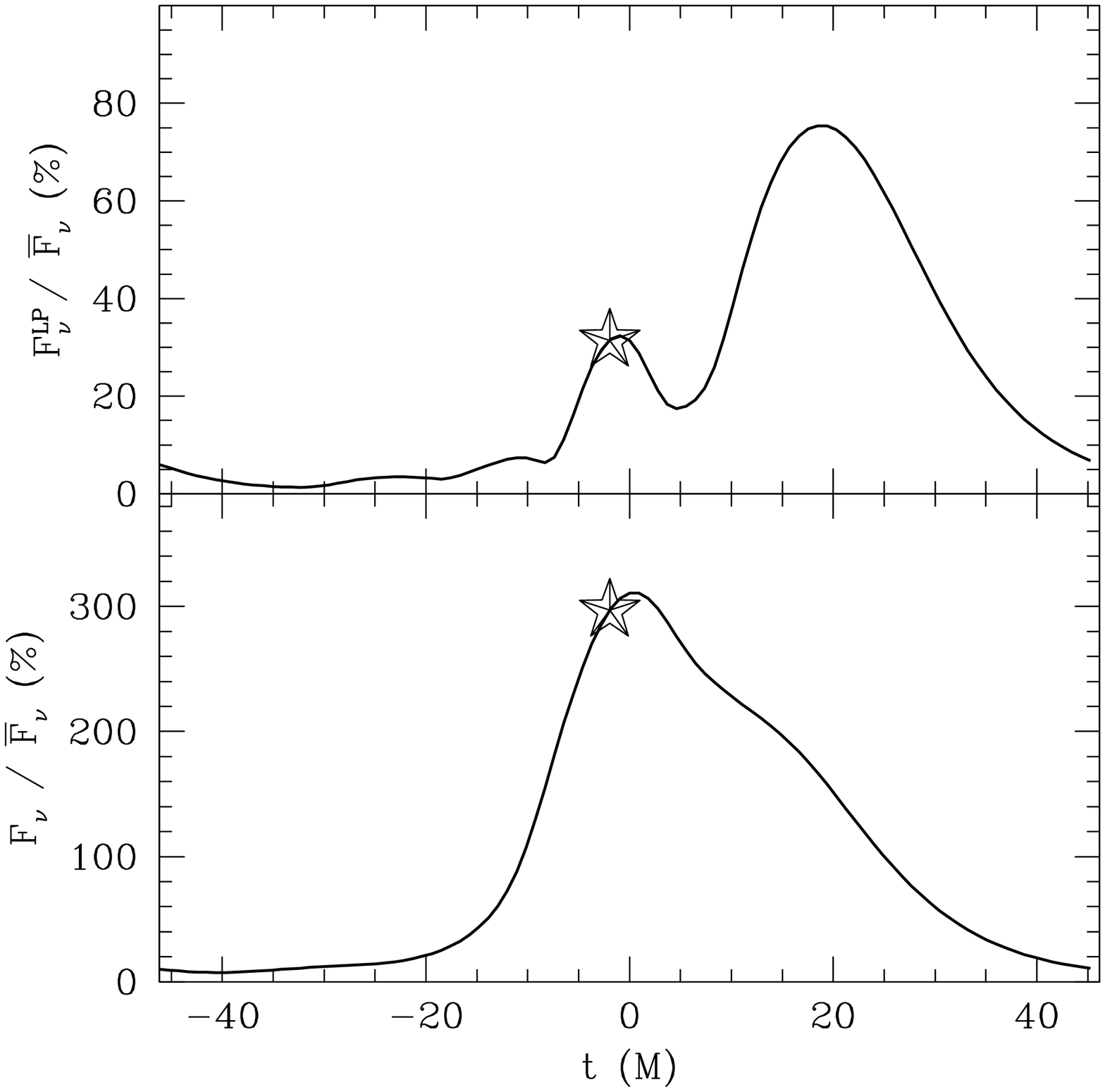}
\caption{\label{fig:lightcurve}Typical light curves for the total
  flux, polarized flux and position angle.  The star indicates the
  time at which the image in Figure \ref{fig:spot_image} was
  produced.  Note the features in the light curve that are due to the
  higher order images.}
\end{minipage} 
\end{figure}

\begin{figure}[t]
\begin{minipage}{0.5\columnwidth}
\includegraphics[width=\columnwidth]{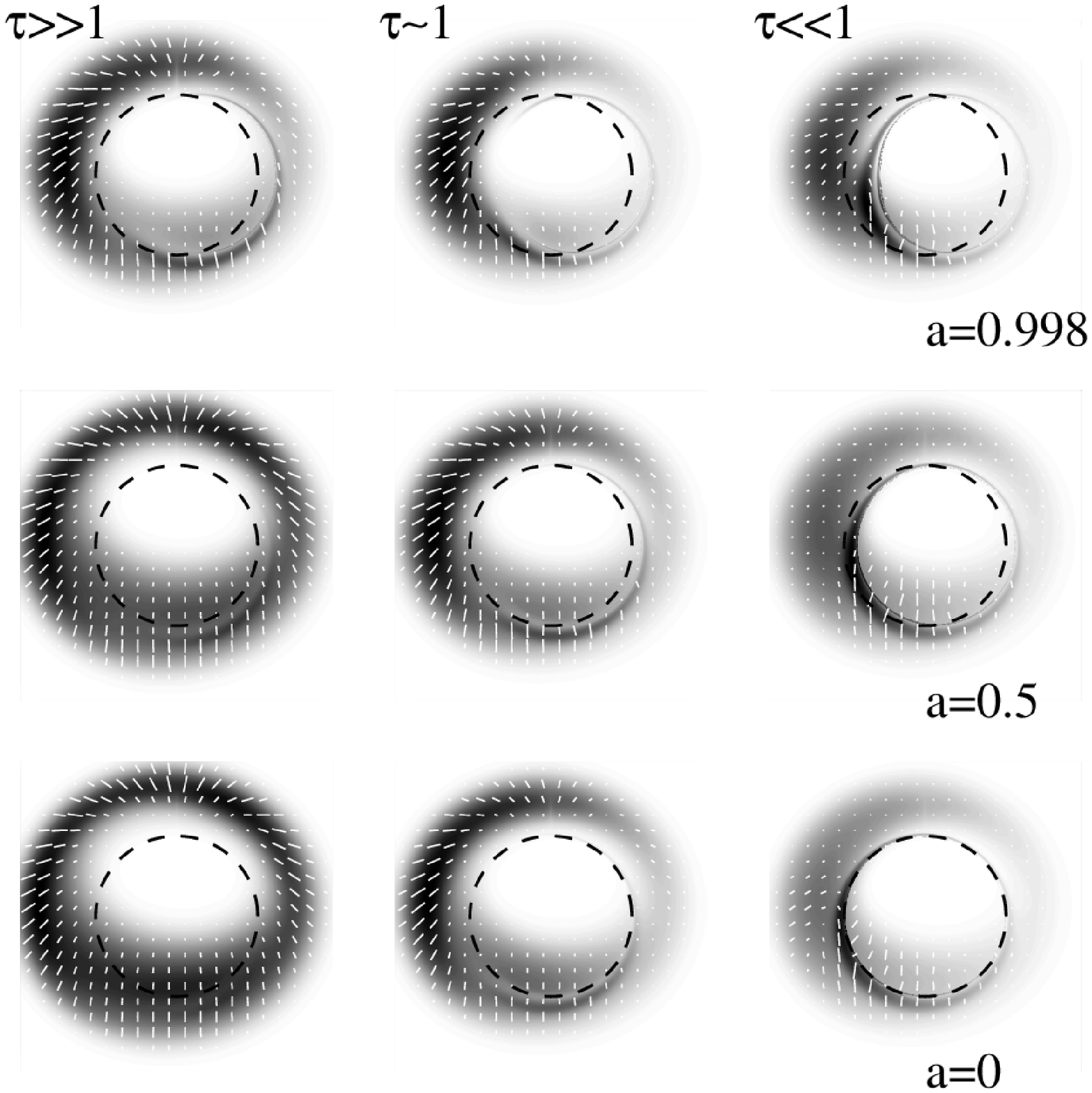}
\caption{\label{fig:averaged_images}Orbit averaged {\em background
    subtracted} images of a hot spot orbiting at a non-rotating,
    moderately rotating, and maximally rotating black hole (bottom to
    top) for low, moderate and high optical depth (left to right).
    Overlayed on the image are polarization ticks in white and the
    size of the silhouette of a non-rotating black hole (the black
    dashed circle).}
\end{minipage}
\hspace{0.05\columnwidth}
\begin{minipage}{0.5\columnwidth}
\includegraphics[width=\columnwidth]{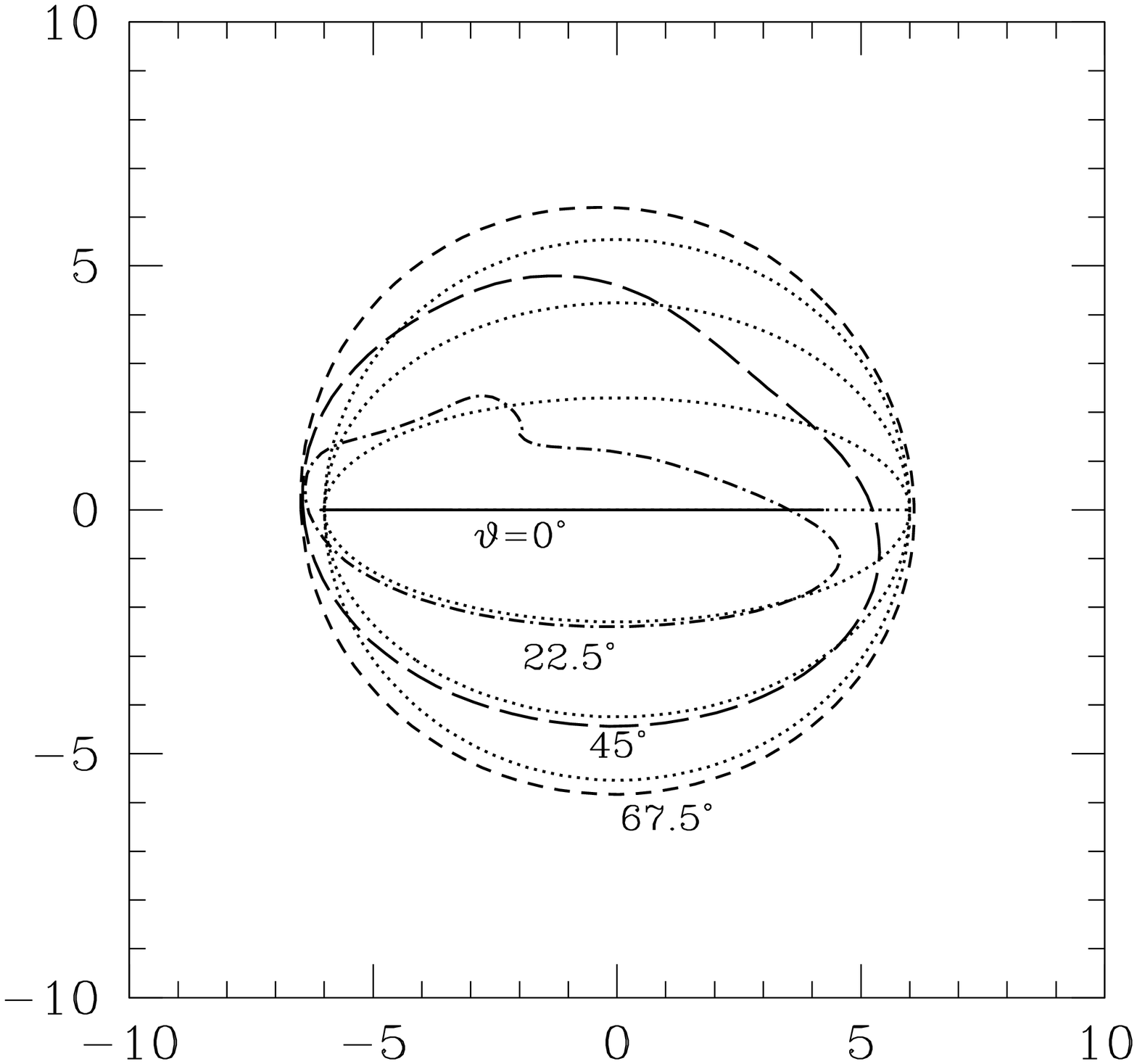}
\caption{\label{fig:centroid_paths}The path of the image centroid for
  a hot spot orbiting a non-rotating black hole at the ISCO for
  various inclinations, $\vartheta$.  For references, the projected
  Newtonian orbits are shown by the dotted lines.}
\end{minipage} 
\end{figure}

Figure \ref{fig:spot_image} shows the instantaneous image of a
spherically symmetric, synchrotron emitting hot
spot (see \cite{Brod-Loeb:06} for details regarding the spot model and the
  generation of the image).
Immediately visible is the fact that two, albeit merging, images of
the hot spot are present.  Within each image the polarization angles
are roughly uniform.  However, the polarization angles of the two
images are significantly different.  The result is that strong lensing
is typically associated with a decrease in the polarized flux.  This
is indeed apparent in the light curve shown in Figure
\ref{fig:centroid_paths}.  Thus, the polarized flux is sensitive to
the presence of higher order images.  Since the light rays which
produce these images are more strongly lensed than those that produce
the primary, and consequently spend more time in the strongly gravitating
region, they are more sensitive the spacetime.  Therefore, as
mentioned earlier, the polarization evolution is diagnostic of the
black hole parameters.

Images of the hot spots are likely to be averaged over many orbits.
Figure \ref{fig:averaged_images} shows {\em background subtracted}
orbit-averaged hot-spot images for the same cases presented in Figure
\ref{fig:silhouettes}.  As with the accretion disk, the hot-spot
images are generally asymmetric, becoming more so as the optical depth
decreases.  However, in this case the images form sets of rings, one
for each hot-spot image.  As spin increases, these rings are displaced, with those
formed by higher order images naturally being displaced further.
Therefore, if the hot-spot parameters (orbital parameters, size,
density and spectral index) are known, then the spin of the black hole
may be determined by measuring the {\em relative} shift between the
primary and secondary images.  Since this involves relative
astrometry, this is an ideal problem for submillimeter VLBI.

In addition to the submillimeter imaging, the proposed GRAVITY
instrument at the very-large telescope interferometer promises to provide
$\sim10\,\muas$ phase-referenced astrometry on few minute timescales
in the NIR \cite{Boug_etal:06,Eise_etal:05}.  Therefore, it will be
possible to track the image centroid path taken by a hot spot.  A seen in
Figure \ref{fig:centroid_paths}, deviations due to strong lensing
do occur.  These are, in general, functions of the black hole spin.
However, the dominant effect is due to the orbital parameters
(inclination and radius).  Hence, at the least, monitoring the
centroid motion provides a method by which to determine the hot-spot
orbit.

\begin{figure}[h]
\includegraphics[width=18pc]{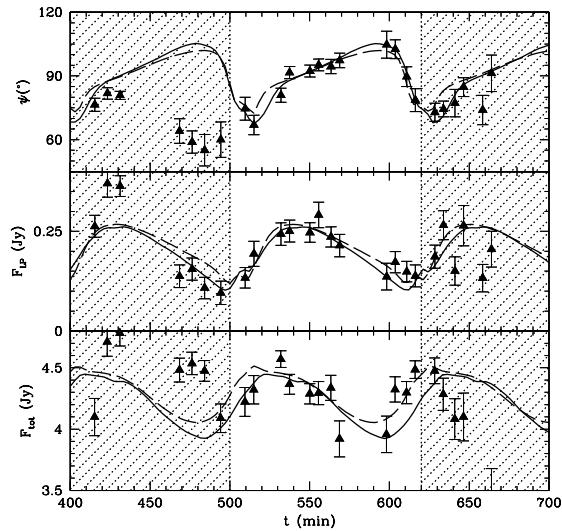}
\hspace{2pc}
\begin{minipage}[b]{18pc}
\caption{\label{fig:preliminary_fits}A preliminary fit of a
  total flux, polarized flux and position angle light curves for a
  submillimeter flare \cite{Marr:06}.  The fit was performed in the
  unhatched region only.  For the dashed line the entire region was
  used.  For the solid line only minutes from $500$ to $600$ were
  used.  In both cases the reduced-$\chi^2$ is of order unity.}
\end{minipage}
\end{figure}

Despite its simplicity, the toy hot spot model we have employed here
fits actual flare intensity and polarization light curves remarkably
well.  Two fits to submillimeter observations are shown in Figure
\ref{fig:preliminary_fits}, differing in the time region over which
the fits were attempted.  Clearly, the flare is evolving on orbital
timescales, with the period decreasing as expected for infall.  This
was not taken into account in the light curve modeling, and thus the
fit was restricted to a single period.  Nevertheless, the hot spot
model is capable of fitting the total intensity, polarized flux and
position angle simultaneously, with a reduce-$\chi^2$ of order unity.  While
a claim to have measured the flare orbit and/or black hole spin would
be premature, the fact that even the simplest model fits implies that
such hot spots are indeed simple.

Combining spectopolarimetric observations, imaging, and the centroid
paths, it is possible in principle to determine simultaneously the
hot-spot orbit and size in physical units, and thus the black hole
mass and spin \cite{Brod-Loeb:06,Brod-Loeb:05}.  Because they are
compact, the rays connecting the hot spot and an observer at infinity
sample only a small portion of the spacetime around the black hole.
Thus, the measured mass and spin are actually indicative of the
spacetime in a small region.  Observations of multiple spots,
presumably at different orbital radii, will consequently provide a way
in which to measure the mass and spin at a number of distinct points
near the black hole.  These may then be compared to the general
relatavistic prediction that the spacetime is fully described by a
mass and spin alone, thus providing a way in which to quantitatively
test the Kerr metric.

\section{Conclusions} \label{C}
Despite their putative size, imaging the black hole horizons at
submillimeter wavelengths is now technologically feasible.  In
addition, phase-referenced astrometry will be available in the NIR in
the coming decade.  These capabilities will allow, for the first time,
direct observations of gravity in the strongly non-linear regime.

Observations of the size, shape and location of the silhouette cast by
the black hole on the surrounding accretion flow may be used in
principle to determine its mass and spin.  However, in practice this
is unlikely to be simple due to uncertainties in accretion physics.
Nevertheless, at the very least it will provide a means to learn about
accretion onto compact objects.

Of more interest will likely be observations of hot spots.  Due to
their compact and dynamical nature their images contain more
information than those of the underlying quiescent accretion flow.  It
appears to be possible, even with the simplest models, to constrain
the hot spot parameters (\eg, orbit, size and spectral index).  If
this is indeed the case, then hot-spot observations will provide a
method for quantitatively testing general relativity.

\ack A.E.B. gratefully acknowledges the support of an ITC Fellowship from
Harvard College Observatory. A. L. was supported in part by NASA grants NAG
5-1329 and NNG05GH54G and by the Clark/Cooke fund of Harvard University.

\bibliographystyle{iopart_nu.bst}

\providecommand{\newblock}{}

\end{document}